\documentclass[12pt,preprint]{aastex}
\title{Dead Zones and the Origin of Planetary Masses} 
\author
{Soko Matsumura \& Ralph E. Pudritz,\\
\normalsize{Department of Physics \& Astronomy, McMaster University,}\\
\normalsize{Hamilton, ON, L8S 4M1, Canada}\\
\normalsize{soko@physics.mcmaster.ca, pudritz@physics.mcmaster.ca}
}
\begin{document}
%
%--- Abstract ---
\begin{abstract}
Protoplanets accrete material from their natal protostellar disks
until they are sufficiently massive to open a gap in the face of the
disk's viscosity that arises from the magneto-rotational instability (MRI).
By computing the ionization structure within observationally
well-constrained disk models, we demonstrate that poorly ionized, 
low viscosity ``dead zones'' stretch out to 12 AU within typical disks.
We find that planets of terrestrial mass robustly form within the dead zones while
massive Jovian planets form beyond.
%We also show that dead zones can significantly slow planetary
%migration.
Dead zones will also halt the rapid migration of planets into their
central stars.
%Finally, we apply our model to the distribution of planetary masses in our own solar system.
Finally, we argue that the gravitational scattering of low mass
planets formed in the dead zone, to larger radii by a rapidly accreting
Jupiter beyond, can explain the distribution of planetary masses in our solar system.
\end{abstract}
\keywords{planetary systems: formation - solar system: formation - planets: general
- accretion disks - MHD - stars: pre-main-sequence}
%
%----------------------------------------------------------------------------
%
% Introduction
%
\section{Introduction}
Current models for planet formation suggest either that Jovian planets
are formed through gas accretion onto cores with \(\sim 10\) Earth
masses that are themselves assembled out of planetesimals
\citep{Mizuno80,Pollack96}, or through gravitational collapse when
protostellar disks become cold and/or dense enough to be locally
gravitationally unstable \citep{Boss97,Mayer02}.
However, independent of how their protoplanetary cores were formed, 
planets must open gaps in their natal disk before they become isolated. 
%The former picture predicts that the formation of a
%Jupiter mass planet at its observed location of 5 AU takes an
%uncomfortably long time; \(\sim 10^6\) years -- about the lifetime of
%gaseous protostellar disks, while the latter has the advantage of a
%shorter formation time of \(\sim 10^3\) years.
%In this Letter, we focus on the fact that planets open gaps in their natal
%disk independently of how their protoplanetary cores were formed.
Ultimately, it is the process of gap-opening that terminates their
growth or at least severely restricts their final masses.
A planet must be sufficiently massive to tidally open a gap in the
face of the disk's viscosity which acts to fill in a forming gap 
\citep{Lin93,Ida04}.

In this Letter, we calculate the gap-opening masses within protostellar 
disk models that are well constrained by the observations and analyzed in
\cite{Matsumura03} (MP03), 
and show that there should be a distinct break in planetary masses 
akin to the difference between terrestrial and Jovian planetary masses.
The reason for this break is that protostellar disks are likely to
contain regions of very low viscosity
-- dead zones -- in which the planetary gap-opening mass is much smaller
than in regions of normal viscosity.
In \S 2, we lay out the theory of gap-opening masses in turbulent disks.
%and apply this to our model in \S3.  
We updated our MP03 disk models to include the effect of a nearby OB
star, additional ionization sources, dust grains, as well as the
turbulence driven in a dead zone by active layers (\S 3) and find that
terrestrial mass planets form within the dead zones, while gas giant
planets form outside of them (\S 4).  
Finally, in \S 5, we apply our results to show that dead zones can
halt the rapid inward migration of protoplanets and that scattering of
low mass planets in the dead zone to larger disk radii by a rapidly
accreting gas giant beyond, could explain the structure of our own
solar system.

\section{Gap-opening masses and disk viscosity}
The minimum gap-opening mass in a fully turbulent region of a protostellar disk,
\(M_{p,turb}\), is attained when the angular momentum transport rate by
a disk's turbulent viscosity becomes equal to that by a planetary tidal force \citep{Lin93}.
It is readily shown that the gap-opening mass depends only on a disk aspect ratio 
\(h/r_{p}\) (h is the local pressure scale height of a disk and
\(r_{p}\) is an orbital radius of the planet) and the turbulent viscosity parameter 
\(\alpha_{{\rm turb}}\).
For the case that the disk's turbulence is absent, we applied the
minimum gap-opening mass \(M_{p,damp}\) obtained by \cite{Rafikov02}.
\cite{Lin93} also calculated a gap-opening mass in an inviscid disk by
assuming that the density waves shock and damp immediately, but this requires a
rather large planetary mass -- the mass inside the Hill radius \(r_{H}=h\). 
\cite{Rafikov02} showed that the density waves excited by a much smaller planet
eventually shock and lead to the gap formation.
%The disk's viscosity may not vanish completely in such a ``dead zone'' however.
%Density waves generated by multiple disk-planet interactions damp to
%deposit angular momentum into the disk and work as an effective disk
%viscosity \citep{Goodman01}.
%The effective viscosity parameter for this case is
%\(\alpha_{{\rm turb}}\sim10^{-4}\) or even smaller, so the corresponding
%viscosity gives a similar gap-opening mass to the inviscid case \(M_{p,damp}\).
The ratio of the gap-opening planetary mass for a region that is fully
turbulent, compared to that for an inviscid disk, can be written as
\begin{equation}
\label{gap12}
\frac{M_{p,turb}}{M_{p,damp}} \geq
2.5 \sqrt{\alpha_{turb}}\left(\frac{h}{r_{p}}\right)^{-23/26} Q^{5/13} \ ,
\end{equation}
in all cases relevant to our disk model; 
where \(Q=\Omega c_{s}/(\pi G \Sigma)\) is the Toomre parameter that
measures the gravitational stability of the disk, \(\Omega\) is the
Keplerian angular velocity, and \(\Sigma\) is the surface mass
density.
%Both gap-opening masses depend on the disk aspect ratio \(h/r_{p}\).
For standard parameters, the critical mass ratio is of the order 
\(M_{p,turb}/M_{p,damp} \sim 100\) which is comparable to the mass
difference between terrestrial and Jovian planets.

A major source of a disk's viscosity is thought to be hydromagnetic
turbulence that is driven by the magneto-rotational instability (MRI)
\citep{Balbus98}.
This mechanism requires good coupling between the partially ionized
gas of the disk and the magnetic field.
Poor coupling, which occurs when this gas is not ionized
sufficiently, leads to the formation of a so-called dead zone wherein
the MRI disk viscosity is effectively zero.
%and hence there is no angular momentum transport.
The spatial extent of a dead zone has been calculated by many authors
\citep{Gammie96,Sano00,Glassgold00,Fromang02,Matsumura03}.
%The dead zone was originally defined by \cite{Gammie96} as the region
%where the MRI instability cannot develop.
Physically, the size of a dead zone can be determined by the condition
that the local growth rate of the MRI turbulence (\(\simeq V_{A}/h\))
becomes smaller than the Ohmic diffusion rate (\(\simeq \eta/h^2\))
where \(\eta\) is the magnetic diffusivity, and \(V_{A} \equiv
B/\sqrt{4 \pi \rho} \sim \alpha^{1/2}_{{\rm turb}} c_{s}\) 
is the Alfv\'{e}n speed (\(\rho\) is the mass density, \(B\)
is the magnetic flux density, and \(c_{s}\) is the sound speed).
%, and \(\alpha_{{\rm turb}}\) is the viscosity parameter). 
Detailed numerical calculations show that this occurs when the ratio
of these two growth rates -- known as the magnetic Reynolds number:
\begin{equation}
\label{rem}
Re_{M} \equiv \frac{V_{A}h}{\eta},
\end{equation}
is less than a critical value of \(Re_{M, {\rm crit}}=10^2-10^4\) \citep{Fleming00}. 
We use \(Re_{M, {\rm crit}}=10^2-10^4\) and \(\alpha_{{\rm turb}}=10^{-3}-1\)
in this paper \footnotemark[1].
Since the magnetic diffusivity \(\eta\) is inversely proportional to
the electron fraction, the MRI turbulence tends to be absent in poorly
ionized regions of the disk.

In the surface layers of a disk, MRI turbulence will always be present and can
drive vertical oscillations into the dead zone below.
This process leads to angular momentum transport whenever the dead zone is not significantly
denser than the active layers: 
\(\Sigma_{{\rm DZ}}/\Sigma_{{\rm AL}} \leq 10\) \citep{Fleming03}, 
where \(\Sigma_{{\rm DZ}}\) and \(\Sigma_{{\rm AL}}\) are the surface
mass density of the dead zone and the active layers respectively.
Therefore, the ``true" dead zone should satisfy both conditions and we
calculated it using a self-consistent model for the disk structure.
%, we calculate how it will be ionized and therefore locate its internal dead zone.

\section{Disk Model}
Most stars are thought to be formed in star clusters such as the Orion
nebula cluster.
A protostellar disk in such an environment will be
irradiated by nearby luminous (\(10^5-10^6 L_{\odot}\)) OB stars in the cluster.
At the outer part of the protostellar disk, the disk heating by the UV
radiation from a central T-Tauri star (TTS) is overwhelmed by the
combined radiation field of the nearby OB stars.
As the outer disk temperature increases, the disk flares more strongly
and therefore planets are expected to have a higher gap-opening mass.
Recently, \cite{Robberto02} (RBP02) improved the
self-consistent model of passive protostellar disks of \cite{CG97} by
taking into account the effect of an external OB star.
They showed that the disk aspect ratio changes significantly.
We used their model to calculate the vertical structure of a disk
around a TTS which is exposed to OB stars \footnotemark[2].

Following RBP02 and without loss of generality, we assume that a disk
is oriented face-on with respect to an O star which is located at a
typical distance of 0.1 pc from the disk and has the luminosity of 
\(L_{s}=6 \times 10^{38} \ {\rm erg \ s^{-1}}\) and a Str\"{o}mgren radius 
of \(R_{{\rm HII}}=10^{18} \ {\rm cm}\).
In all of our models, we adopt the disk surface mass density at 1 AU of 
\(\Sigma_{0}=10^3\) or \(10^4 \ {\rm g \ cm^{-2}}\) which is equivalent
to a disk mass inside 100 AU of \(\sim 0.01\) or \(0.1 M_{\odot}\)
respectively (\(M_{\odot}\) is the mass of the Sun).
These disk masses are typical for T Tauri stars \citep{Hartmann98,Kitamura02}.

The electron fraction that controls the magnetic diffusivity of disks,
is determined by the balance between ionization and recombination.
For the ionization sources, we include X-rays from the central star
and a nearby O star, cosmic rays, radioactive elements as well as the
thermal collisions of alkali atoms.
Among those, X-rays from the central star and cosmic rays are the two
major ionizing sources of protostellar disks (e.g. MP03). 
Protostellar X-rays may be generated within large loops of
stellar magnetic field that result from reconnection during
magnetospheric accretion, as shown in time-dependent calculations \citep[e.g.][]{Hayashi96}.
We assume that magnetic loops extend out to distances \((r,z)=(2R_{*}, 2R_{*})\)
from the stellar surface and adopt a typical T Tauri star's X-ray
luminosity \(L_{x}=10^{30} \ {\rm erg \ s^{-1}}\) and
temperature \(kT_{x}=2 \ {\rm keV}\) (Feigelson, private communication).
We also assume that cosmic rays propagate along field lines and that the strength of
cosmic rays is constant across the disk surface.
Both ionization rates as well as the uniform ionization by radioactive
elements: \(\xi_{{\rm RA}}=6.9 \times 10^{-23} \ {\rm s}^{-1}\) are
calculated as in MP03.
In this paper, we also include the thermal ionization of alkali metals (potassium)
due to heating by the central star following \cite{Fromang02}.
They showed that there is a magnetically active zone at the innermost
radii (\(r\lesssim0.1\) AU).
We find, however, that the thermal ionization effect is killed if we
take account of the recombination on grains (see below).
Our dead zone stretches from the inner disk radius to a few tens of AU. 
%This effect makes thick active layers at the smallest radii, leading to
%the stimulated MRI inside a dead zone at the innermost region.
%and found that their effect is eliminated by considering the MRI
%penetration from active layers into a dead zone.
We also find that the X-ray ionization by a fiducial nearby O star
with the X-ray luminosity \(L_{x}=10^{34} \ {\rm erg \ s^{-1}}\) and
temperature \(kT_{x}=2 \ {\rm keV}\) is too weak to affect the
ionization structure of the disk.

For recombination processes, we considered the reactions among
electrons, molecular ions, metals and grains.  
At disk density \(n\geq n_{{\rm crit}} \sim 10^{12} \ {\rm cm^{-3}}\),
grains are very effective at reducing the charge in disks and hence
increasing the diffusivity \(\eta\).
Were it not for the stimulation of turbulence in the body of the disk from the
envelope, our detailed calculations, which followed the method by
\cite{Umebayashi90} and \cite{Fromang02}, show that very extensive
dead zones -- out to 16 -- 29 AU -- are to be expected \footnotemark[3].
%while other reactions are calculated following \cite{Fromang02} as in MP03.
However, this ``stimulated'' turbulence limits the extent of a dead
zone to smaller radii, typically 12 -- 25 AU for our two fiducial disk
column densities.
  
\section{Results}
\begin{figure}[t]
\unitlength1cm
\begin{minipage}[t]{8cm}
\begin{picture}(8,5)
\scalebox{0.3}{
\includegraphics{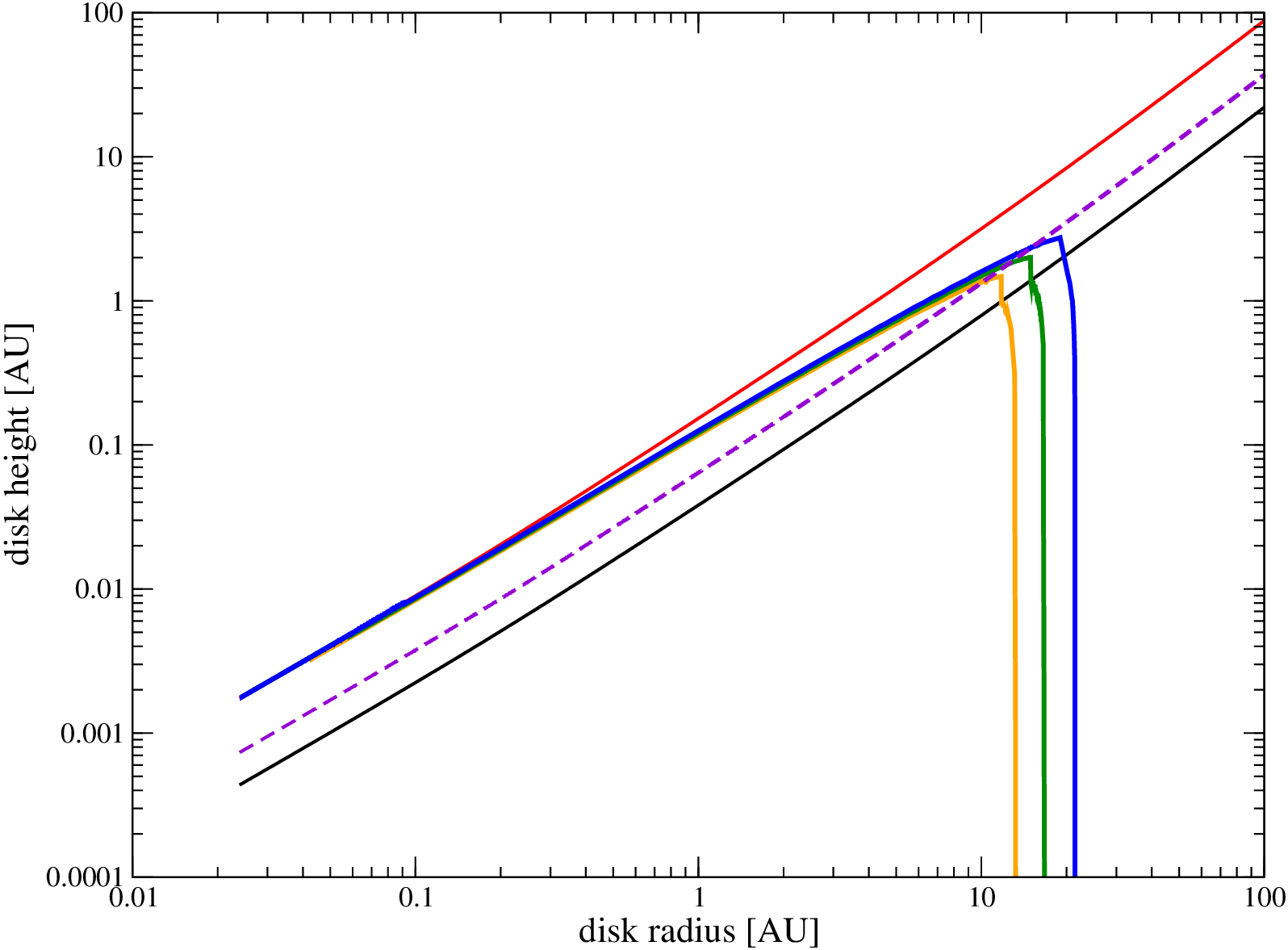}}
\end{picture}
\caption[dead1]{Dead zones calculated for the standard disk model with an
external star. We used \(\Sigma_{0}=10^3 \ {\rm g \ cm^{-2}}\),
\(L_{x}=10^{30} {\rm erg \ s^{-1}}\), \(kT_{x}=2\) keV, and
\(Re_{M}=10^3\). The uppermost line shows the surface disk height; the
lowermost line shows the pressure scale height while three curves show
the dead zone boundaries for \(\alpha=\) 0.1, 0.01, and
0.001 from inside to outside. The dashed line is where 
\(\Sigma_{{\rm below}}/\Sigma_{{\rm above}}=10\). \label{dead1}}
\end{minipage}
\hspace{0.5cm}
\begin{minipage}[t]{8cm}
\begin{picture}(8,7)
\scalebox{0.3}{
\includegraphics{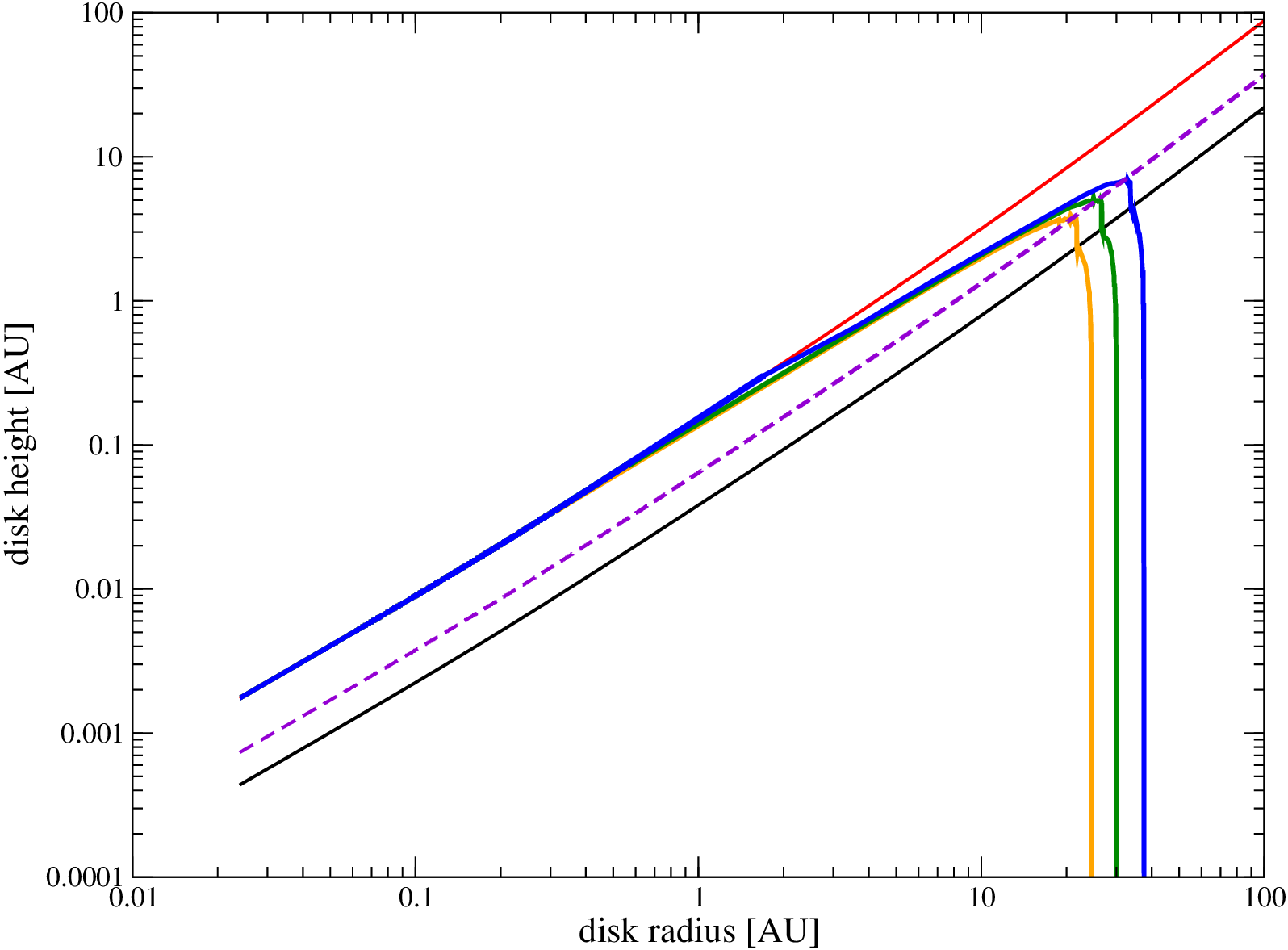}}
\end{picture}
\caption[dead2]{Dead zones calculated for a heavier disk of
\(\Sigma_{0}=10^4\ {\rm g \ cm^{-2}}\) with an external star. For the
explanation of each line, see Fig. \ref{dead1}. \label{dead2}}
\end{minipage}
\end{figure}

We show the spatial structure of disks as well as their internal dead
zones in Fig. \ref{dead1} and \ref{dead2} which correspond to disk
column densities of \(\Sigma_{0}=10^3\) and \(10^4 \ {\rm g \ cm^{-2}}\) respectively.
Three curves show the disk height \(z_{{\rm rem}}\) where the 
magnetic Reynolds number \(Re_{M}\) reaches its critical value
\(10^3\) for \(\alpha_{{\rm turb}}=0.1, \ 0.01\), and \(0.001\) from
inside to outside.
Also shown is the disk height \(z_{{\rm surf}}\) where the surface
mass density ratio of below and above it is equal to 10;
\(\Sigma_{{\rm below}}/\Sigma_{{\rm above}}=10\) (a dashed line).
We define the dead zone as the region where \(z\leq z_{{\rm rem}}\)
and \(z_{{\rm rem}}>z_{{\rm surf}}\) -- the intersection of
\(z_{{\rm rem}}\) and \(z_{{\rm surf}}\) marks the outer dead
zone radius
As the magnetic field becomes stronger (the parameter \(\alpha_{\rm
turb}\) becomes larger), the dead zone becomes smaller.
For a standard disk (\(\Sigma_{0}=10^3 \ {\rm g \ cm^{-2}}\)) with \((Re_{M},\alpha_{{\rm
turb}})=(10^3,0.01)\), we find that the dead zone stretches from the
inner edge of a disk to 12 AU and encompasses the terrestrial planet
region in our solar system (\(0.3 - 2\) AU).
For a moderately heavy disk (\(\Sigma_{0}=10^4 \ {\rm g \ cm^{-2}}\)),
the dead zone radius becomes 25 AU.
These results agree well with previous works \citep[e.g.][]{Sano00}.
It may be possible that cosmic rays do not ionize disks because they
are swept away by disk winds. 
In this case, disk ionization is determined mainly by X-rays from the
central star and an external star.
%solely by the stellar X-rays. 
The dead zone radii then become slightly larger (14 AU for 
\(\Sigma_{0}=10^3 \ {\rm g \ cm^{-2}}\) and 36 AU for 
\(\Sigma_{0}=10^4 \ {\rm g \ cm^{-2}}\)).
%Note also that the surface layers are completely turbulent at \(\leq
%0.03 - 0.07\) AU due to good coupling provided by the thermal ionization
%close to the central star.
%We applied this constraint to our disk models 
%and found that our conclusions are not affected 
%- this density ratio is typically \(\sim 10^2 - 10^4\).

%We determined the inner and outer dead zone
%radii as in Fig. \ref{mass1} and \ref{mass2}
%by computing the ionization structure as well as
%by including only regions with sufficiently high
%column density contrast.
In Fig. \ref{mass1} and \ref{mass2}, we show the calculated
gap-opening masses of planets for disks with \(\Sigma_{0}=10^3\) and \(10^4
\ {\rm g \ cm^{-2}}\) respectively.
Both disks are gravitationally stable (we find \(Q_{{\rm min}} > 3\) in
our \(\Sigma_{0}=10^4 \ {\rm g \ cm^{-2}}\) model) so that planet
formation by gravitational instability should not occur in these disk models.
The upper parallel horizontal lines show the gap-opening masses with various
viscous parameters while the bottom solid horizontal line shows
the gap-opening masses in an inviscid disk.
Vertical lines with crosses indicate the extent of a dead zone.
%Also plotted is a Hill sphere mass which is defined by \cite{Lin93} as
%a gap-opening mass in an inviscid disk.
A thick solid line shows the fiducial gap-opening mass throughout the disk.

In the core accretion scenario, the minimum gap-opening mass inside
the dead zone is dictated by wave damping (the lowermost line), while
outside, the minimum gap-opening mass is determined
by the strength of the MRI viscosity and hence the strength of the
magnetic field (the value of \(\alpha_{{\rm turb}}\)).
The predicted planetary mass therefore makes a distinct jump upwards to
a value determined by the value of \(\alpha_{{\rm turb}}\) outside
the dead zone.
Our fiducial cases, shown in Fig. \ref{mass1} and \ref{mass2} with a
thick solid line, corresponds to \(\alpha_{{\rm turb}}=0.01\) outside
the dead zone.
This \(\alpha_{{\rm turb}}\) value is suggested by observations of
protostellar disks on scales \(\sim 10 -  100\) AU \citep{Hartmann98,Kitamura02}.
Along this line in Fig. \ref{mass1}, we find the gap-opening mass
becomes equal to an Earth mass \(M_{E}\) at \(\sim 0.7\) AU and a
Jupiter mass \(M_{J}\) at \(\sim 17\) AU. 
Our results show that Jupiter or more massive gas giant planets
must form outside the dead zones while terrestrial and
perhaps even ice giant planets (see below) are likely to
form within them.
%
%In the gravitational instability scenario, planets are expected to
%form by consuming most of the gas inside the Hill sphere whose radius is as large as the
%pressure scale height.
%Inside the Hill radius, the planetary tidal force on disk gas
%exceeds the stellar tidal force.
%The mass corresponding to the local Hill sphere is given by
%\(M_{p}/M_{*}=3(h/r)^3\) which is plotted in Fig. \ref{mass1} and
%\ref{mass2}. 
%As can be seen in the figures, the gap-opening mass predicted by the Hill
%sphere is actually less than the mass predicted for disk's
%viscosity exceeding \(\alpha_{{\rm turb}} \geq\) few \(0.01\).
%Thus MRI viscosity will continue to drive the growth of planets formed
%by gravitational instability to larger values than that expected by
%simple Hill spheres if the disk has a strong magnetic field.
%For example, in a disk with \(\Sigma_{0}=10^3\) or \(10^4 \ {\rm g \ cm^{-2}}\) and
%the viscosity parameter of \(\alpha_{{\rm turb}}=0.1\), a forming planet of
%mass \(\sim 1-2 \ M_{J}\) at \(10-20\) AU would grow up to \(\sim 2-4 \ M_{J}\) 
%before a gap is opened and the planetary accretion ceases.
%These masses are comparable to the results from the simulations of
%gravitationally unstable disks \citep{Mayer02}.  
%However, \(\alpha_{{\rm turb}}>0.01\) may be too large to be explained
%by the recent observations \citep{Hartmann98,Kitamura02}.

\section{An integrated picture for solar system formation}
Our results show that massive planets in our fiducial disk models form beyond 12 AU.
This further supports the need of planet migration as an explanation of the
observed exosolar systems.
The presence of a dead zone may solve a nagging problem of migration
theory -- there is no general mechanism of halting a planet's
migration into the central star.
The standard migration picture consists of two types of migration.
When a protoplanet is not very massive, it migrates through the
disk without opening a gap (type I).
As a protoplanet gains a sufficient mass, it opens a gap in the disk
and subsequently migrates with the disk on a viscous timescale (type II).
The type I migration is roughly two orders of magnitudes faster than the type II
migration \citep{Ward97,Terquem03}.
This means that in an inviscid region like a dead zone, a migrating
planet will be stalled as soon as it opens a gap \citep{Chiang02}.
Since the presence of finite dead zones is a robust feature of the
protostellar disks, they may act as natural barriers that prevent the
rapid loss of planets into their central stars.
%It is likely that the disk has some other sources of viscosity, like
%damping of density waves, inside the dead zone.
%Even in such a case, planets may still be saved due to the smaller
%gap-opening masses inside a dead zone -- migrating into the dead zone
%may switch a rapid, type I migration to a slower, type II migration.
%This slows the speed of migration significantly so that
%the disk gas may dissipate before the planet plunges into the star.
%The time scale for the type I migration is \(\tau_{{\rm I}}({\rm
%year}) \sim 10^8(M_{E}/M_{p})\Sigma \sqrt{{\rm AU}/r} \times 10^2 (h/r)^2\) while 
%that for the type II migration is \(\tau_{{\rm II}}({\rm year}) \sim 
%1/(3\alpha_{{\rm turb}}\Omega)(r/h)^2\) \citep{Ward97,Terquem03}.
%For a standard disk (\(\Sigma_{0}=10^3 \ {\rm g \ cm^{-2}}\)) with
%\(\alpha_{{\rm turb}}=0.01\), the type I migration timescale becomes
%\(\sim 1500\) years for a Jupiter mass planet, \(\sim 5000\) years for a
%Saturn mass planet, and \(\sim 50000\) years for a 10 Earth mass planet at
%the outer radius of the dead zone (\(\sim 8\) AU).
%The type II migration timescale with the viscosity of \(\alpha_{{\rm
%turb}}=10^{-4}-10^{-5}\) will be \(\sim 2 \times 10^6\) years and
%\(\sim 2 \times 10^7\) years (about a typical disk lifetime) at \(\sim
%8\) AU respectively. 

The planetary masses predicted by our disk model (and others) 
increase with disk radius.
In our solar system, this is not observed - the lower mass ice giants
Uranus and Neptune are found at larger radii (19.2 and 30.1 AU respectively).
One scenario which can explain this mass sequence is the
photoevaporation of the disk that reduces the surface mass density
significantly  beyond \(\sim 15\) AU within \(\sim 10^7\) years \citep{Adams04}
but this may be too long for currently accepted disk lifetimes. 
    
\begin{figure}[p]
\unitlength1cm
\begin{minipage}[t]{8cm}
\begin{picture}(8,5)
\scalebox{0.3}{
\includegraphics{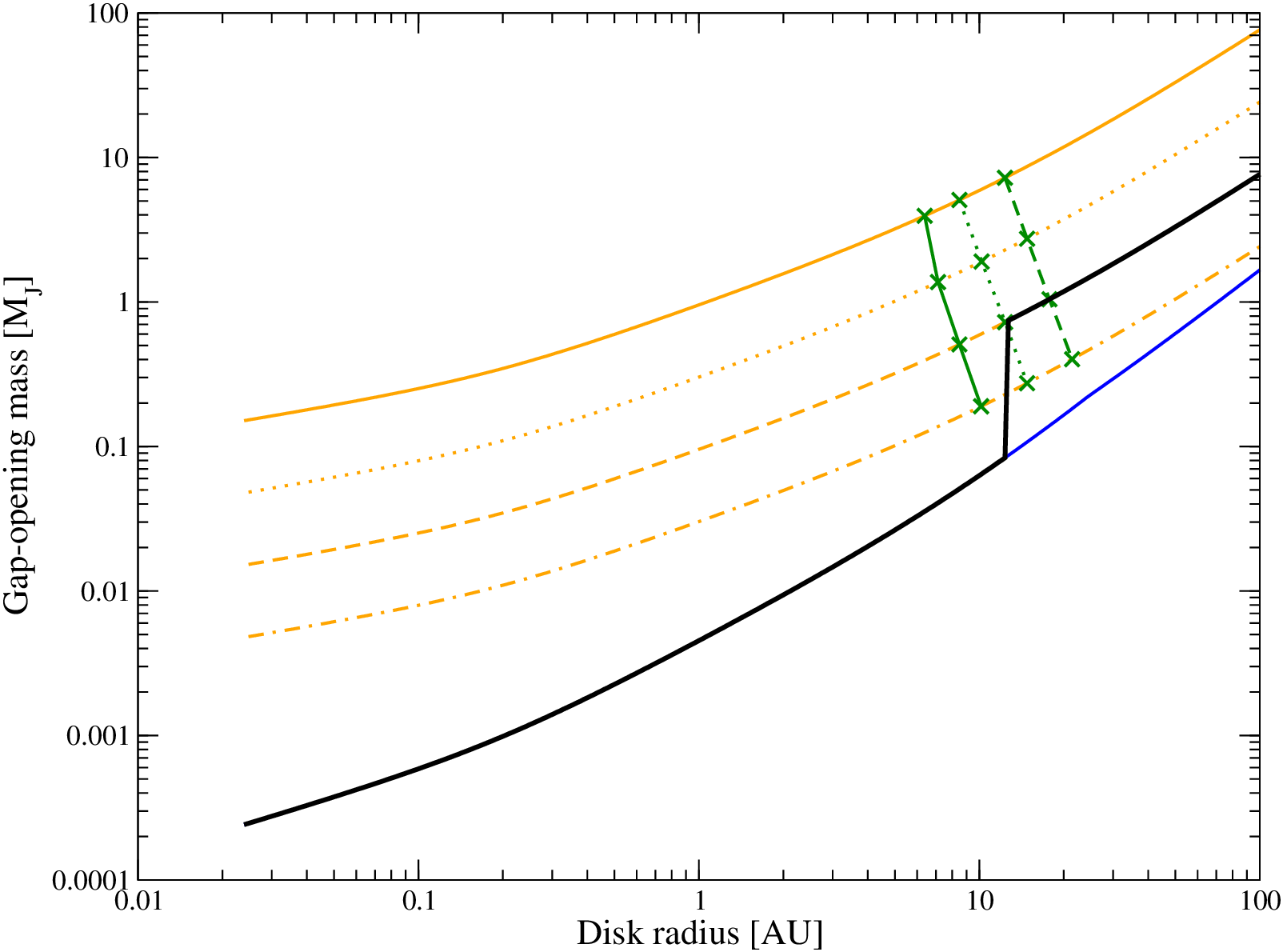}}
\end{picture}
\caption[gap1]{Gap-opening masses for a disk with the surface mass
density, at 1 AU, of \(\Sigma_{0}=10^3 \ {\rm g \ cm^{-2}}\) (or 
$M_{d} \sim 0.01M_{\odot}$).  The lowermost line shows the gap-opening
mass for the region with no MRI viscosity while the upper parallel
lines show the gap-opening masses for a different strength of magnetic
field: solid line is for \(\alpha_{{\rm turb}}=1\), dotted line is for
0.1, dashed-line is for 0.01, and dot-dashed line is for 0.001.  The
dead zone for the magnetic Reynolds number:\(Re_{M}=10^2\) is inside a solid
line with crosses, that for \(10^3\) is inside a dotted line with
crosses, and that for \(10^4\) is inside a dashed line with crosses.  
The thick solid line is our fiducial minimum gap-opening mass throughout the
disk for MRI ``viscosity'', \(\alpha_{{\rm turb}}=0.01\). \label{mass1}}
\end{minipage}
\hspace{0.5cm}
\begin{minipage}[t]{8cm}
\begin{picture}(8,7)
\scalebox{0.3}{
\includegraphics{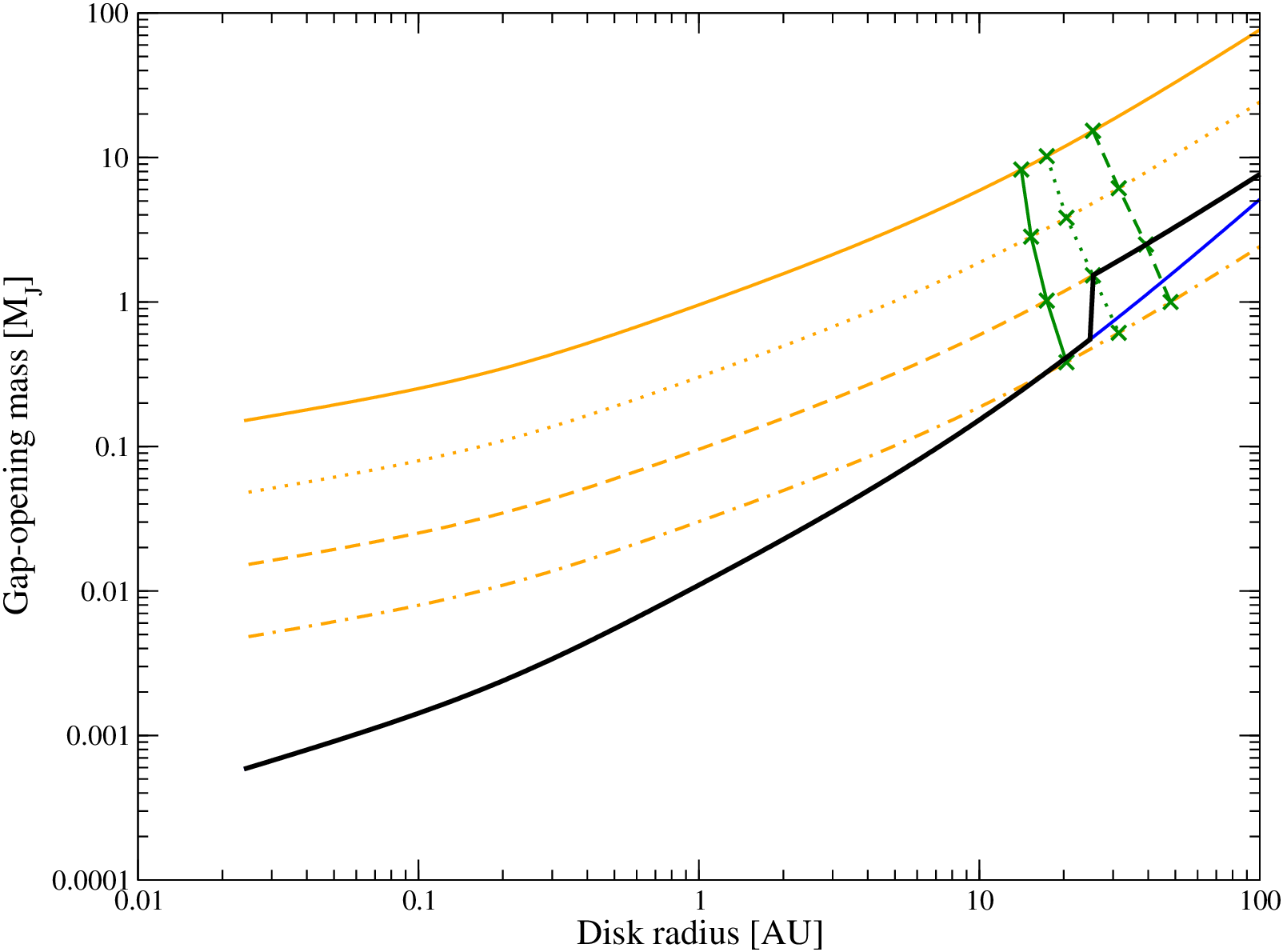}}
\end{picture}
\caption[gap2]{Gap opening masses for a disk with the surface mass
density at 1 AU of \(\Sigma_{0}=10^4 \ {\rm g \ cm^{-2}}\) (or 
$M_{d} \sim 0.1M_{\odot}$).  For the explanation of each line, see
Fig. \ref{mass1}. \label{mass2}}
\end{minipage}
\end{figure}

We suggest an alternative - the gravitational scattering of lower
mass protoplanets from within the dead zone to much larger radii
by a gas giant located just outside of it.
Numerical experiments \citep{Thommes99} have shown that a rapidly
accreting Jupiter can scatter a more slowly growing protoplanetary
core on an interior orbit as the former's Hill sphere expands.
The scattered low mass body will ultimately circularize its orbit
by dynamical friction at sufficiently large disk radii
to be decoupled from the scatterer.
This scenario has the distinct advantage of building the cores of ice
giants much faster because it happens in the inner region of the disk.
This process may occur naturally in our disk model.
For example, in Fig. \ref{mass1} along the curve 
\((Re_{M}, \alpha_{{\rm turb}})=(10^3,0.01)\), we find that a \(0.74
M_J\) planet is formed just outside the dead zone at \(r=12\) AU, while 
the maximum mass just inside the dead zone is \(0.08 M_J\) -
which is roughly equal to the mass of an ice giant planet.
The inner core(s) would be scattered when the disk gas becomes sufficiently
tenuous so that the eccentricity can be excited.

An inevitable consequence of a dead zone is that material from the
well-coupled region beyond its outer edge will accumulate at this
interface.
The increasing column density of such an annulus may push the dead
zone outward in radius, and may even become gravitationally unstable.
%Meanwhile, the disk will rapidly lose its inner, well-coupled zone,
%opening an inner hole in the disk.
We are currently investigating these time-dependent problems.
%The scattered core will rapidly accrete an icy mantle in the outer regions. 

In conclusion, we have calculated planetary gap-opening masses in standard
mass (\(\sim 0.01 \ M_{\odot}\)) and moderately heavy (\(\sim 0.1 \
M_{\odot}\)) disk models which are exposed to a nearby O star.
With widely accepted values of \((Re_{M},\alpha_{{\rm turb}}) = (10^3,
0.01)\), we have found that the dead zone stretches out to 12 AU for
the standard mass disk.
We have shown that there is a distinct and model-independent range of
planetary masses within the dead zone compared to the well-coupled zone
beyond.
With \((Re_{M}, \alpha_{{\rm turb}})=(10^3,0.01)\), this corresponds to terrestrial
mass planets (\(2.4 \times 10^{-4} - 8.3 \times 10^{-2} \
M_{J}\)) vs Jovian mass planets (\(0.74 - 7.6 \ M_{J}\)).
The robust nature of our results leads us to conclude that dead zones
are typical in protostellar disks and may play a central role in
determining the masses of planetary families as well as their fates. 

We thank Eric Feigelson, David Hollenbach, Doug Lin, Edward Thommes and James Wadsley
for stimulating discussions and an anonymous referee for a very useful
report.
S. M. is supported by a SHARCNET Graduate Fellowship while R. E. P. is
supported by a grant from the National Science and Engineering
Research Council of Canada (NSERC).
\bibliography{REF}

\begin{thebibliography}{27}
\expandafter\ifx\csname natexlab\endcsname\relax\def\natexlab#1{#1}\fi

\bibitem[{{Adams} {et~al.}(2004){Adams}, {Hollenbach}, {Laughlin}, \&
  {Gorti}}]{Adams04}
{Adams}, F.~C., {Hollenbach}, D., {Laughlin}, G., \& {Gorti}, U. 2004, ApJ,
  611, 360

\bibitem[{{Balbus} \& {Hawley}(1998)}]{Balbus98}
{Balbus}, S.~A. \& {Hawley}, J.~F. 1998, Reviews of Modern Physics, 70, 1

\bibitem[{{Boss}(1997)}]{Boss97}
{Boss}, A.~P. 1997, Science, 276, 1836

\bibitem[{{Chiang} {et~al.}(2002){Chiang}, {Fischer}, \& {Thommes}}]{Chiang02}
{Chiang}, E.~I., {Fischer}, D., \& {Thommes}, E. 2002, ApJL, 564, L105

\bibitem[{{Chiang} \& {Goldreich}(1997)}]{CG97}
{Chiang}, E.~I. \& {Goldreich}, P. 1997, ApJ, 490, 368

\bibitem[{{Chiang} {et~al.}(2001){Chiang}, {Joung}, {Creech-Eakman}, {Qi},
  {Kessler}, {Blake}, \& {van Dishoeck}}]{Chiang01}
{Chiang}, E.~I., {Joung}, M.~K., {Creech-Eakman}, M.~J., {Qi}, C., {Kessler},
  J.~E., {Blake}, G.~A., \& {van Dishoeck}, E.~F. 2001, ApJ, 547, 1077

\bibitem[{{Fleming} \& {Stone}(2003)}]{Fleming03}
{Fleming}, T. \& {Stone}, J.~M. 2003, ApJ, 585, 908

\bibitem[{{Fleming} {et~al.}(2000){Fleming}, {Stone}, \& {Hawley}}]{Fleming00}
{Fleming}, T.~P., {Stone}, J.~M., \& {Hawley}, J.~F. 2000, ApJ, 530, 464

\bibitem[{{Fromang} {et~al.}(2002){Fromang}, {Terquem}, \&
  {Balbus}}]{Fromang02}
{Fromang}, S.~., {Terquem}, C., \& {Balbus}, S.~A. 2002, MNRAS, 329, 18

\bibitem[{{Gammie}(1996)}]{Gammie96}
{Gammie}, C.~F. 1996, ApJ, 457, 355

\bibitem[{{Glassgold} {et~al.}(2000){Glassgold}, {Feigelson}, \&
  {Montmerle}}]{Glassgold00}
{Glassgold}, A.~E., {Feigelson}, E.~D., \& {Montmerle}, T. 2000, {Protostars
  and Planets IV} (The University of Arizona Press, 2000)

\bibitem[{{Hartmann} {et~al.}(1998){Hartmann}, {Calvet}, {Gullbring}, \&
  {D'Alessio}}]{Hartmann98}
{Hartmann}, L., {Calvet}, N., {Gullbring}, E., \& {D'Alessio}, P. 1998, ApJ,
  495, 385

\bibitem[{{Hayashi} {et~al.}(1996){Hayashi}, {Shibata}, \&
  {Matsumoto}}]{Hayashi96}
{Hayashi}, M.~R., {Shibata}, K., \& {Matsumoto}, R. 1996, ApJL, 468, L37+

\bibitem[{{Ida} \& {Lin}(2004)}]{Ida04}
{Ida}, S. \& {Lin}, D.~N.~C. 2004, ApJ, 604, 388

\bibitem[{{Kitamura} {et~al.}(2002){Kitamura}, {Momose}, {Yokogawa}, {Kawabe},
  {Tamura}, \& {Ida}}]{Kitamura02}
{Kitamura}, Y., {Momose}, M., {Yokogawa}, S., {Kawabe}, R., {Tamura}, M., \&
  {Ida}, S. 2002, ApJ, 581, 357

\bibitem[{{Lin} \& {Papaloizou}(1993)}]{Lin93}
{Lin}, D.~N.~C. \& {Papaloizou}, J.~C.~B. 1993, {Protostars and Planets III}
  (The University of Arizona Press)

\bibitem[{{Matsumura} \& {Pudritz}(2003)}]{Matsumura03}
{Matsumura}, S. \& {Pudritz}, R.~E. 2003, ApJ, 598, 645

\bibitem[{{Mayer} {et~al.}(2002){Mayer}, {Quinn}, {Wadsley}, \&
  {Stadel}}]{Mayer02}
{Mayer}, L., {Quinn}, T., {Wadsley}, J., \& {Stadel}, J. 2002, Science, 298,
  1756

\bibitem[{{Mizuno}(1980)}]{Mizuno80}
{Mizuno}, H. 1980, Progress of Theoretical Physics, 64, 544

\bibitem[{{Pollack} {et~al.}(1996){Pollack}, {Hubickyj}, {Bodenheimer},
  {Lissauer}, {Podolak}, \& {Greenzweig}}]{Pollack96}
{Pollack}, J.~B., {Hubickyj}, O., {Bodenheimer}, P., {Lissauer}, J.~J.,
  {Podolak}, M., \& {Greenzweig}, Y. 1996, Icarus, 124, 62

\bibitem[{{Rafikov}(2002)}]{Rafikov02}
{Rafikov}, R.~R. 2002, ApJ, 572, 566

\bibitem[{{Robberto} {et~al.}(2002){Robberto}, {Beckwith}, \&
  {Panagia}}]{Robberto02}
{Robberto}, M., {Beckwith}, S.~V.~W., \& {Panagia}, N. 2002, ApJ, 578, 897

\bibitem[{{Sano} {et~al.}(2000){Sano}, {Miyama}, {Umebayashi}, \&
  {Nakano}}]{Sano00}
{Sano}, T., {Miyama}, S.~M., {Umebayashi}, T., \& {Nakano}, T. 2000, ApJ, 543,
  486

\bibitem[{{Terquem}(2003)}]{Terquem03}
{Terquem}, C.~E.~J.~M.~L.~J. 2003, preprint (astro-ph/0309175)

\bibitem[{{Thommes} {et~al.}(1999){Thommes}, {Duncan}, \&
  {Levison}}]{Thommes99}
{Thommes}, E.~W., {Duncan}, M.~J., \& {Levison}, H.~F. 1999, Nature, 402, 635

\bibitem[{{Umebayashi} \& {Nakano}(1990)}]{Umebayashi90}
{Umebayashi}, T. \& {Nakano}, T. 1990, MNRAS, 243, 103

\bibitem[{{Ward}(1997)}]{Ward97}
{Ward}, W.~R. 1997, Icarus, 126, 261

\end{thebibliography}
\bibliographystyle{apj}
\footnotetext[1]{Recent numerical simulations by \cite{Fleming00} 
                 defined the magnetic Reynolds number as
                 \(\acute{Re_{M}}=c_{s}h/\eta\) and determined the critical value in
                 the range of \(10^2 - 10^4\). 
                 In our definition of \(Re_{M}=\alpha^{1/2}c_{s}h/\eta\), 
                 \((Re_{M},\alpha_{{\rm turb}})=(10^2,0.01)\) and \((10^3,0.01)\)
                 correspond to their \(\acute{Re_{M}}=10^3\) and \(10^4\) respectively.}
\footnotetext[2]{We compared three different grain models which give the emissivity
                 difference and therefore the temperature difference
                 of up to \(\sim 40\) \%.
                 Two of them are RBP02's models of Type I and Type II grains whose
                 radius is \(0.1 \ {\rm \mu m}\) and \(0.02 \ {\rm \mu m}\)
                 respectively.
                 The other is the disk model by \cite{Chiang01} with an
                 external stellar radiation.
                 This grain model gives the lowest temperature of all. 
                 We chose an intermediate model which uses the optical depth of 
                 \(\tau_{\bot}=0\) and the type I grains of RBP02.}
\footnotetext[3]{To be consistent with the disk model, we assumed all
                 grains have the same radius \(0.1 \ {\rm \mu m}\).  
                 Our electron fraction for a solar nebula model is
                 about 2 orders of magnitude larger than the one
                 obtained by \cite{Sano00} because we ignore grains of
                 charge \(\pm 2e\).  With their electron fraction and
                 our choice of \(Re_{M}=10^3\), the dead zone would be
                 only slightly larger.}
\end{document}